\documentclass[3p,times]{elsarticle}

\usepackage{ecrc}

\usepackage{epstopdf}

\volume{00}

\firstpage{1}

\journalname{Nuclear Physics A}

\runauth{Author1 et al.}

\jid{nupha}

\jnltitlelogo{Nuclear Physics A}

\usepackage{graphicx}
\usepackage{amsmath,amssymb}
\begin{document}

\begin{frontmatter}



\title{Searches for $p_{\rm T}$ dependent fluctuations of flow angle and magnitude in Pb--Pb and p--Pb collisions}

%% Single author (and collaboration) - please insert
\author{You Zhou (for the ALICE\fnref{col1} Collaboration)}
\fntext[col1] {A list of members of the ALICE Collaboration and acknowledgements can be found at the end of this issue.}
\address{Nikhef, Science Park 105, 1098 XG Amsterdam, The Netherlands\\ 
Utrecht University, P.O.Box 80000, 3508 TA Utrecht, The Netherlands\\
you.zhou@cern.ch
}

%% For multiple authors, replace the above by:

%\author[label1]{Author1}
%\author[label2]{Author2}

%\address[label1]{Address 1}
%\address[label2]{Address 2}

\begin{abstract}

Anisotropic azimuthal correlations are used to probe the properties and the evolution of the system created in heavy-ion collisions. 
Two-particle azimuthal correlations were used in the searches of $p_{\rm T}$ dependent fluctuations of flow angle and magnitude, measured with the ALICE detector.
The comparison of hydrodynamic calculations with measurements will also be presented in this proceedings.

%In addition, the factorization of the two-particle Fourier harmonics $V_{n\Delta}$ for different values of $n$ into single-particle azimuthal anisotropies $v_{n}$, together with the comparison to hydrodynamic model calculations, will be discussed in both Pb--Pb and p--Pb collisions.

%A template for preparing contributions to the proceedings of Quark Matter 2014. The file should be compiled with {\em pdflatex}. Figures can be pdf or eps, as illustrated in Fig.~\ref{fig:generic}. If the conversion eps $\to$ pdf does not work automatically on your system, you can convert eps files to pdf using a tool like {\em epstopdf}. For special options see\\ 

\end{abstract}


\end{frontmatter}

% \linenumbers


\section{Introduction}
\label{intro}

The primary goal of ultra relativistic heavy-ion collisions is to understand the properties of the quark-gluon plasma (QGP), a new state of matter whose existence under extreme conditions is predicted by quantum chromodynamics. 
An important experimental observable toward this goal is the study of anisotropic flow using a Fourier expansion of the azimuthal anisotropy~\cite{Voloshin:1994mz},
\begin{equation}
E\frac{d^{3}N}{d^{3}{\bf p}} = \frac{1}{2\pi} \frac{d^{2}N}{p_{\rm T}dp_{\rm T}d\eta}(1+2\sum_{n=1}^{\infty}v_{n} \cos[n(\varphi - \Psi_{n})])
\end{equation}
\label{eq:FourierExp}
where $E$ is the energy, ${\bf p}$ the momentum, $p_{\rm T}$ the transverse momentum, $\varphi$ the azimuthal angle, $\eta$ the pseudorapidity of the particle. %and  the $n-$order participant plane (or called symmetry plane) is defined as $\Psi_{n}$
The $v_{n}$ coefficients and $\Psi_{n}$ represent the magnitude and angle (symmetry plane) of the $n^{th}$-order harmonic, respectively. 
The elliptic flow $v_{2}$, triangular flow $v_{3}$, quadrangular flow $v_{4}$, and pentagonal flow $v_{5}$ have been measured at the CERN Large Hadron Collider (LHC). It provides compelling evidence that strongly interacting matter appears to behave like an almost perfect fluid~\cite{Heinz:2013th}. 
Recent hydrodynamic simulations predict the $p_{\rm T}$ dependent fluctuations of flow angle and magnitude.
%that flow angle $\Psi_{n}$ depends on $p_{\rm T}$, $\Psi_{n}(p_{\rm T})$ fluctuate around the event average symmetry plane $\Psi_{n}$~\cite{Heinz:2013bua,Gardim:2012im}. 
Two new observables, $v_{n}\{2\}/v_{n}[2]$ and $r_{n}$ were proposed in hydrodynamic calculations~~\cite{Heinz:2013bua,Gardim:2012im}. They are used to check whether the factorization of two-particle correlations into the product of single particle anisotropy harmonics is valid, which probes the $p_{\rm T}$ dependent flow angle and magnitude fluctuations. 

In this proceeding, we present the measurements of $v_{2}\{2\}/v_{2}[2]$ and factorization ratio $r_{n}$ in Pb--Pb collisions at $\sqrt{s_{\rm NN}} = $ 2.76 TeV and p--Pb collisions at $\sqrt{s_{\rm NN}} = $ 5.02 TeV in ALICE.

%Precise anisotropic flow measurements is crucial for extracting $\eta/s$ and investigating the properties of the created hot and dense matter in heavy-ion collisions. If there is a possible $p_{\rm T}$ dependent fluctuations of flow angle and magnitude, the previous anisotropic flow measurements as well as the extracted $\eta/s$ and initial state condition will be biased due to such effects. Thus, it is critical to probe and constrain the $p_{\rm T}$ dependent fluctuations of flow angle and magnitude in heavy-ion experiments.

\section{Analysis Details}
The data sample collected by ALICE in the first Pb--Pb run and p--Pb run at the Large Hadron Collider were used in this analysis. For more details of ALICE detector, refer to~\cite{Aamodt:2008zz}. About 16 million Pb--Pb events and 97 million p--Pb events were recorded with a minimum-bias trigger, based on signals from two VZERO detectors (-3.7$\textless \eta \textless$-1.7 for VZERO-C and 2.8$\textless \eta \textless$5.1 for VZERO-A) and on the Silicon Pixel Detector. The VZERO detectors were also used for the determination of the collision centrality in Pb--Pb collisions while VZERO-A was taken for the determination of the multiplicity classes in p--Pb collisions. Charged particles are reconstructed using the Inner Tracking System and the Time Projection Chamber with full azimuthal coverage for pseudo-rapidity range $|\eta|\textless$0.8. 
The two-particle correlations are measured using the Q-Cumulant method~\cite{Bilandzic:2010jr} with a pseudorapidity gap $|\Delta \eta| > $ 0.8, which is expected to suppress the non-flow effects (azimuthal correlations not related to the symmetry plane) as much as possible but still allow good statistical precision within the TPC acceptance.

\section{Results}

%\newpage
%{\bf  A generic figure, in pdf and eps, is shown in Fig.~\ref{fig:generic}.}
%\vspace*{1cm}

\begin{figure}[h]
\begin{center}
\includegraphics*[width=13.5cm]{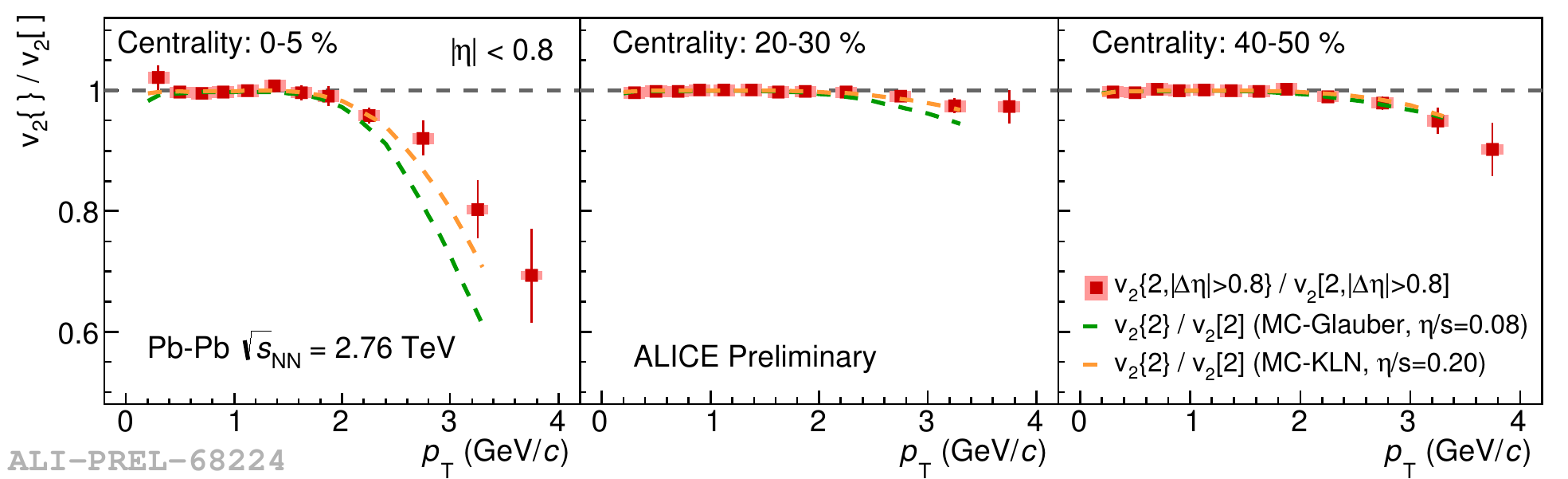}
\caption{
The ratio $v_{2}\{2, |\Delta\eta| > 0.8\}/v_{2}[2, |\Delta\eta| > 0.8]$ for various centralities of Pb--Pb collisions at $\sqrt{s_{\rm NN}} = 2.76$~TeV. The error bars correspond to statistical uncertainties, while the shaded color bands correspond to the systematic uncertainties.
Hydrodynamic calculations with MC-Glauber initial condition and $\eta/s =$ 0.08 and with MC-KLN initial condition and $\eta/s =$ 0.20  are shown in green and orange lines.
}
\label{fig:f1}
\end{center}
\end{figure}

\begin{figure}[h]
\begin{center}
\includegraphics*[width=13.5cm]{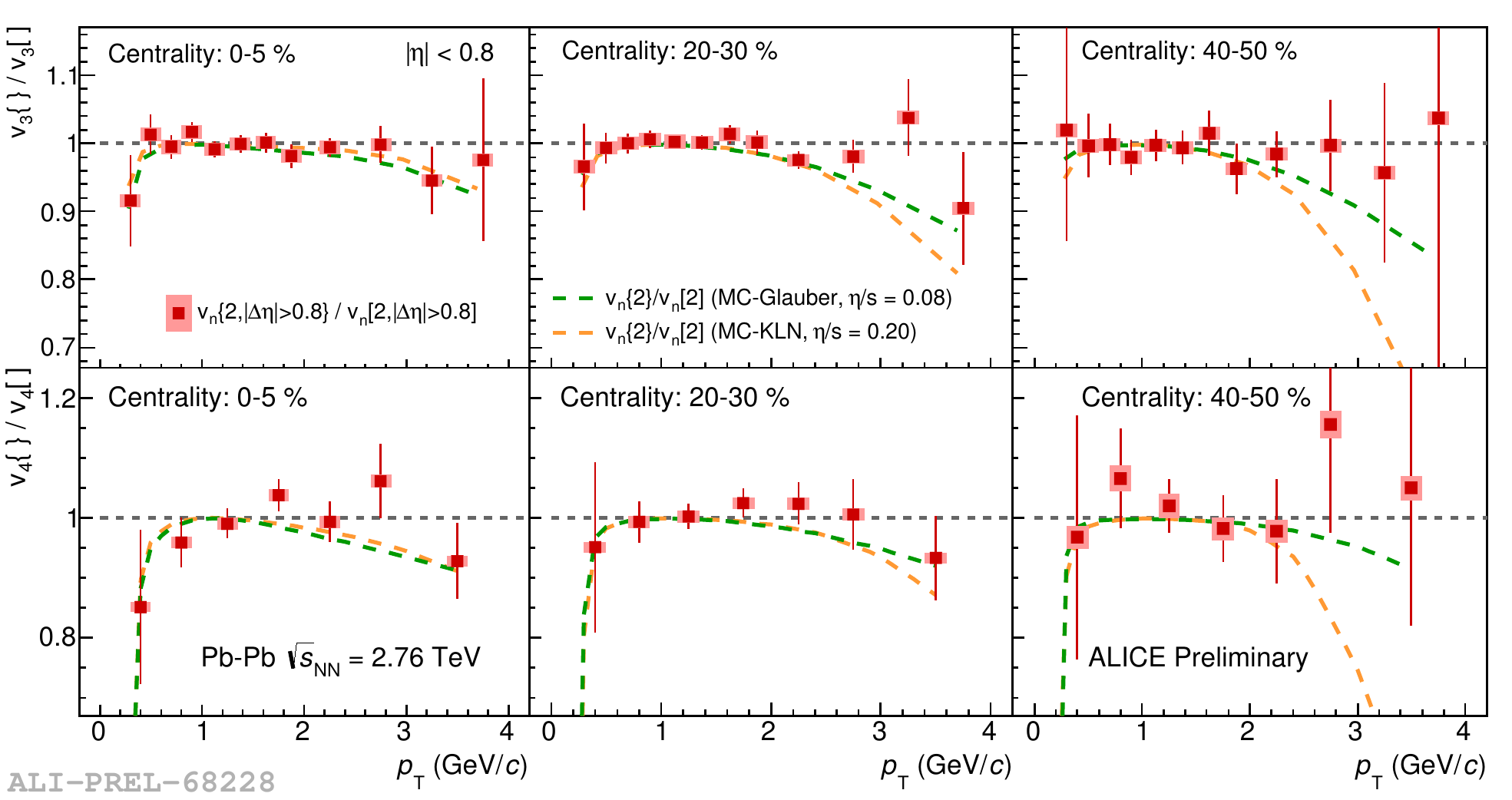}
\caption{
The ratio $v_{3}\{2, |\Delta\eta| > 0.8\}/v_{3}[2, |\Delta\eta| > 0.8]$ for various centralities of Pb--Pb collisions at $\sqrt{s_{\rm NN}} = 2.76$~TeV. 
%Hydrodynamic calculations with MC-Glauber initial condition and $\eta/s =$ 0.08 and with MC-KLN initial condition and $\eta/s =$ 0.20 are shown in  dash green and dash orange curves.
}
\label{fig:f2}
\end{center}
\end{figure}

Figure~\ref{fig:f1} presents the ratio $v_{2}\{2, |\Delta\eta|> 0.8\}/v_{2}[2, |\Delta\eta| > 0.8]$ as a function of $p_{\rm T}$ in different centrality classes. 
It is observed that the ratio is consistent with unity up to $p_{\rm T} \approx$ 2 GeV/$c$ in most central collisions and up to $p_{\rm T} \approx$ 3 GeV/$c$ in the non-central collisions. 
The deviations from unity become weaker but occur at higher $p_{\rm T}$ range toward peripheral collisions.
The result indicates that if $p_{\rm T}$ dependent flow angle ($\Psi_{2}$) and magnitude ($v_{2}$) fluctuations exist, such effect is within 10$\%$ in non-central collisions in the presented $p_{\rm T}$ range.
The comparison to hydrodynamic calculations shows that both calculations overestimate the deviation of $v_{2}\{2, |\Delta\eta|> 0.8\}/v_{2}[2, |\Delta\eta| > 0.8]$ in most central collisions, and the data is better described by the one with MC-KLN initial condition and $\eta/s =$ 0.20.

%our data prefer descriptions of hydrodynamic calculations with MC-KLN initial condition and $\eta/s =$ 0.20 to MC-Glauber initial condition and $\eta/s =$ 0.08. It indicates that the formal calculation might not only generate reasonable $v_{2}$, but also the $p_{\rm T}$ dependent fluctuations of flow angle ($\Psi_{2}$) and magnitude ($v_{2}$).

The ratios of $v_{3}\{2, |\Delta\eta| > 0.8\}/v_{3}[2, |\Delta\eta| > 0.8]$ and $v_{4}\{2, |\Delta\eta| > 0.8\}/v_{4}[2, |\Delta\eta| > 0.8]$ together with various hydrodynamic calculations are shown in Fig.~\ref{fig:f2}. It is found that %the ratio of both $v_{3}\{2, |\Delta\eta| > 0.8\}/v_{3}[2, |\Delta\eta| > 0.8]$ and $v_{4}\{2, |\Delta\eta| > 0.8\}/v_{4}[2, |\Delta\eta| > 0.8]$ 
both ratios agree with unity over a wider $p_{\rm T}$ range than $v_{2}\{2, |\Delta\eta| > 0.8\}/v_{2}[2, |\Delta\eta| > 0.8]$.
The $p_{\rm T}$ dependent fluctuations of flow angle $\Psi_{3}$ (and $\Psi_{4}$) and magnitude $v_{3}$ (and $v_{4}$) are not significant in the presented $p_{\rm T}$ range.
%Even though the hydrodynamic calculations with MC-Glauber and MC-KLN initial conditions can't reproduce the magnitude of $v_{3}\{2\}$ and $v_{3}[2]$, it is still interesting to check if they can generate reasonable $p_{\rm T}$ dependent fluctuations of flow angle ($\Psi_{3}$) and magnitude ($v_{3}$).
%Very rough agreements are observed between data and hydrodynamic calculations, 
Meanwhile it's still difficult to conclude which hydrodynamic calculations describe the measurements better, due to limited statistics.

\begin{figure}[h]
\begin{center}
\includegraphics*[width=13.cm]{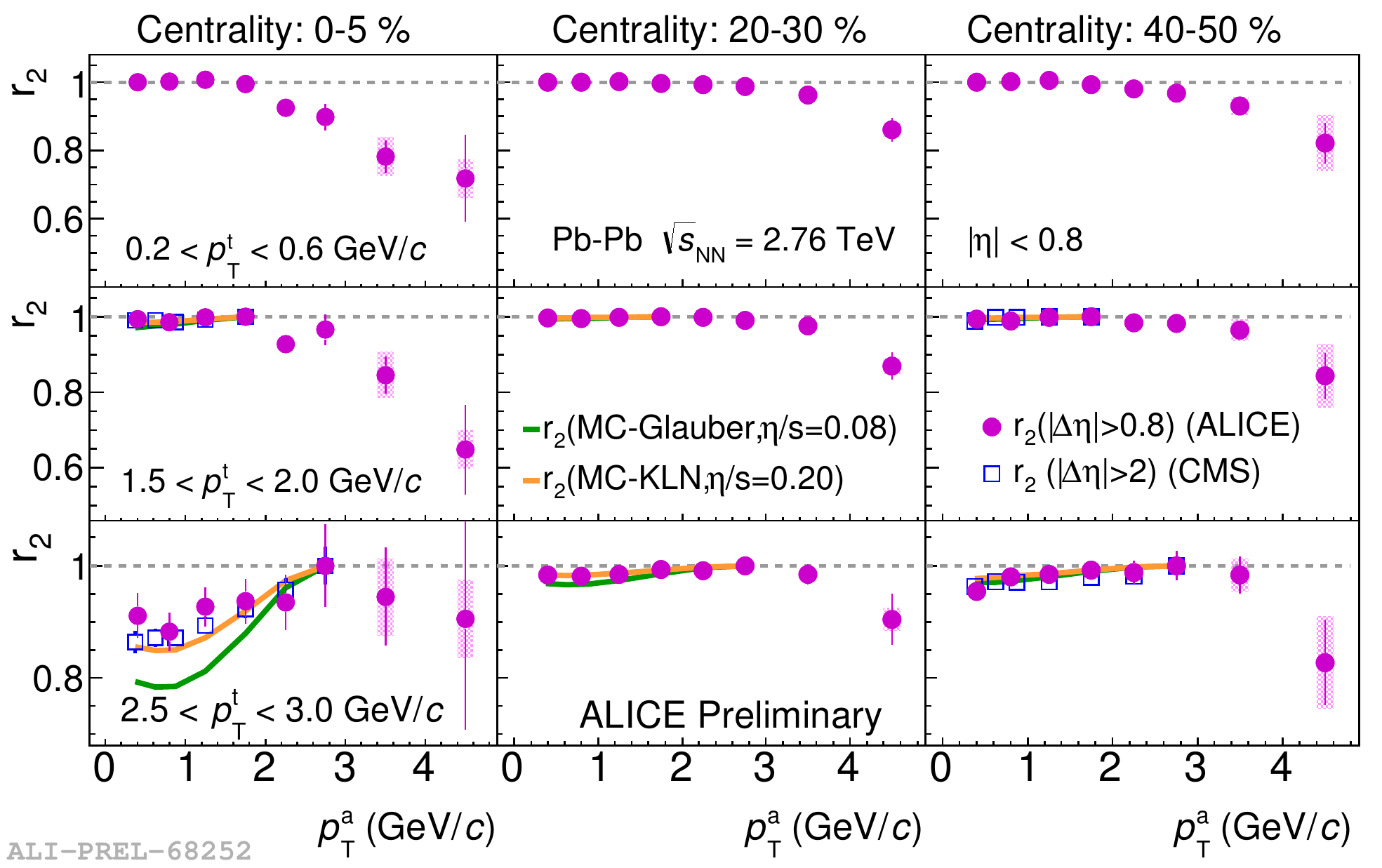}
\caption{
The factorization ratio $r_{2}$, as a function of $p_{\rm T}^{a}$ in bins of $p_{\rm T}^{t}$ for 0-5 $\%$, 20-30 $\%$ and 40-50 $\%$ in Pb--Pb collisions at $\sqrt{s_{_{\rm NN}}} = 2.76$~TeV, is presented by solid circles. CMS measurements~\cite{CMS:2013bza} are presented by open squares, hydrodynamic calculations with MC-Glauber initial condition and $\eta/s =$ 0.08 and with MC-KLN initial condition and $\eta/s =$ 0.20 are shown in green and orange curves.
}
\label{fig:f3}
\end{center}
\end{figure}

Figure~\ref{fig:f3} shows the factorization ratio $r_{2}$ for three $p_{\rm T}$ of triggle particles (denoted as $p_{\rm T}^{t}$), as a function of $p_{\rm T}$ of associate particles (denoted as $p_{\rm T}^{a}$) for centrality classes 0-5 $\%$, 20-30 $\%$ and 40-50 $\%$ in Pb--Pb collisions. 
It significantly deviates from unity as the collisions become more central, this effect becomes stronger with an increase of $|p_{\rm T}^{t} - p_{\rm T}^{a}|$. 
%The previous measurements indicated that the factorization holds approximately for $n >$ 1 and $p_{\rm T}$ below 4 GeV/$c$, while this analysis shows that factorization breaks at lower $p_{\rm T}$ with a more sensitive observable, factorization ratio $r_{2}$.
The deviation is within 10 $\%$ for the lowest $p_{\rm T}^{a}$ in the 0-5 $\%$ for 2.5$\textless~p_{\rm T}^{t}~\textless$ 3.0 GeV/$c$. 
This can be due to $p_{\rm T}$ dependent fluctuations of flow angle ($\Psi_{2}$) and magnitude ($v_{2}$) generated by initial event-by-event geometry fluctuations, and possible remaining non-flow effects. 
Various hydrodynamic calculations are compared to data for the presented centrality classes. Both hydrodynamic calculations with MC-Glauber initial condition and $\eta/s = $ 0.08 and MC-KLN initial condition and $\eta/s =$ 0.20 qualitatively predict the trend of $r_{2}$, while the latter agrees better with data.
In addition, recent CMS measurements~\cite{CMS:2013bza}, which are based on $|\Delta\eta| >$ 2.0 (asymmetry cuts in pseudorapidity), are consistent with our measurements. 
Notice that the ALICE analysis focus on the mid-pseudorapidity range and the selected two correlated particles are always from symmetric sub-events in pseudorapidity, while this is not the case in CMS measurements. The agreement suggest that the possible additional pseudorapidity dependent fluctuations of flow angle ($\Psi_{2}$) and magnitude ($v_{2}$) is not accessible in the presented pseudorapidity range.
The factorization of $r_{3}$ (not shown here) in Pb--Pb collisions is valid over a wider range of $p_{\rm T}^{a}$, $p_{\rm T}^{t}$ and centrality ranges, compared to $r_{2}$. 
The factorization is broken within 10 $\%$ for $p_{\rm T}^{a}$, $p_{\rm T}^{t}$ below 4 GeV/$c$. 
CMS measurements quantitatively agree with our $r_{3}$ measurements, additional pseudorapidity dependent fluctuations of flow angle ($\Psi_{3}$) and magnitude ($v_{3}$) might be negligible in the presented pseudorapidity range.

\begin{figure}[h]
\begin{center}
\includegraphics*[width=14.cm]{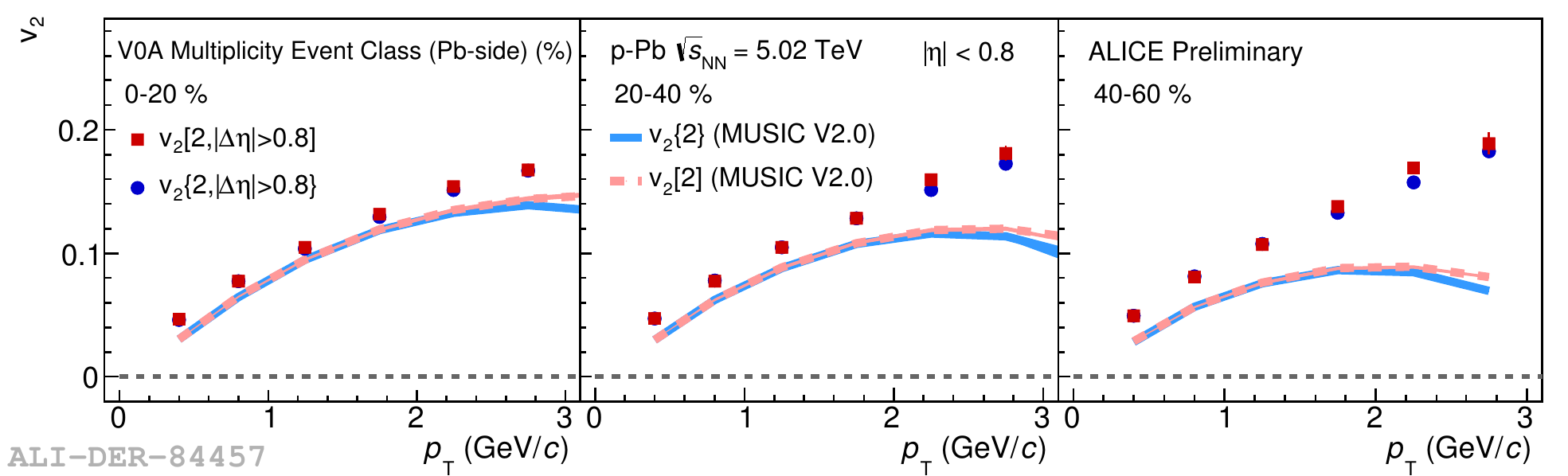}
\caption{
The $v_{2}\{2, |\Delta\eta| > 0.8\}$ and $v_{2}[2, |\Delta\eta| > 0.8]$ for various multiplicity classes of p--Pb collisions at $\sqrt{s_{\rm NN}} = 5.02$ TeV. Hydrodynamic calculations (MUSIC) of $v_{2}\{2\}$ and $v_{2}[2]$ are shown by solid and dash lines.}
\label{fig:f4}
\end{center}
\end{figure}

The $v_{2}\{2, |\Delta\eta| > 0.8\}$ and $v_{2}[2, |\Delta\eta| > 0.8]$ are also measured in p--Pb collisions at $\sqrt{s_{\rm NN}} = 5.02$ TeV in ALICE. It is found that $v_{2}\{2, |\Delta\eta| > 0.8\}$ and $v_{2}[2, |\Delta\eta| > 0.8]$ deviate at $p_{\rm T} \sim$ 3 GeV/$c$ for the presented multiplicity classes. Hydrodynamic calculation~\cite{Kozlov:2014fqa} show a similar trend as a function of $p_{\rm T}$ but shows better agreement in central than in peripheral p--Pb collisions. Due to limited statistics, the $v_{3}\{2, |\Delta\eta| > 0.8\}$ and $v_{3}[2, |\Delta\eta| > 0.8]$ are investigated up to $p_{\rm T} \sim$ 3 GeV/$c$. The difference of this two measurements are not observed for the the presented multiplicity and $p_{\rm T}$ range.

\section{Summary}

In summary, searches of $p_{\rm T}$ dependent flow angle and magnitude fluctuations are performed by measuring $v_{n}\{2\}/v_{n}[2]$ and $r_{n}$. It is found that $v_{2}\{2, |\Delta\eta|> 0.8\}/v_{2}[2, |\Delta\eta| > 0.8]$ is consistent with unity up to $p_{\rm T} \approx$ 2 GeV/$c$ in most central Pb--Pb collisions, the deviation becomes weaker but occur at higher transverse momenta towards peripheral collisions. 
A significant deviation of $v_{3}\{2, |\Delta\eta|> 0.8\}/v_{3}[2, |\Delta\eta| > 0.8]$ and $v_{4}\{2, |\Delta\eta|> 0.8\}/v_{4}[2, |\Delta\eta| > 0.8]$ from unity is not observed. 
The factorization ratio $r_{2}$ significantly deviates from unity as the collisions become more central, such effect becomes stronger as $|p_{\rm T}^{t} - p_{\rm T}^{a}|$ is increasing. The comparison to hydrodynamic calculations shows that the one with MC-KLN initial condition and $\eta/s =$ 0.20 has better agreement to the data than with MC-Glauber initial condition and $\eta/s =$ 0.08, but none of them can quantitatively describe the data. The $v_{n}\{2\}/v_{n}[2]$ has been measured also in p--Pb collisions, it deviates from unity for $p_{\rm T} >$ 3 GeV/$c$ and is under predicted in hydrodynamic calculations.



%% References


%\bibliographystyle{elsarticle-num}



\end{document}